# Characterising transiting exoplanets at long orbital period: lessons learned for PLATO from 10 years of monitoring the HIP41378 system

Alexandre Santerne[1,2], Salomé Grouffal[1] and Stéphane Udry[3]

[1] Univ. Grenoble Alpes, CNRS, IPAG, 38000 Grenoble, France
[2] Aix Marseille Univ, CNRS, CNES, LAM, Marseille, France
[3] Department of Astronomy, University of Geneva

**The upcoming launch of the PLATO mission will open a new area of exoplanetary research by probing transiting exoplanets at long orbital periods around bright stars. These planets will be amenable to follow-up observations, especially with the ESO telescopes as the first PLATO field is in the southern sky. In this paper we listed the lessons we learned by monitoring the bright HIP41378 multi-planetary system over one decade via both radial-velocity and transit observations. These lessons are important for the follow-up strategy of PLATO transiting exoplanets.**

## The HIP41378 system as a testbed for the PLATO mission

Discovered in 2015 by the K2 mission (Vanderburg et al., 2016), the HIP41378 system is a rare target hosting five transiting exoplanets with long orbital period (up to 542 days of period). The host is a bright (V=8.9) late-F dwarf which fundamental parameters have been constrained thanks to asteroseismology (Lund et al., 2019). Among all the planetary systems known to date, this is the only one which combine both a bright, well-characterised host star and several long-period (P>100 days) transiting exoplanets.

The PLATO mission, to be launched early 2027, aims at detecting transiting exoplanets around bright stars for which accurate parameters could be derived through asteroseismology (Rauer et al., 2025). By staring on the same stellar field over a few years, the mission will probe a population of transiting planets at long orbital period. The Kepler mission was also designed to detect such planets (e.g. Kipping et al., 2016), but on much fainter stars[a] (V ~ [14,16]). This was a strong limitation for the ground-based radial-velocity (RV) follow-up of these planets, even for the giant and massive ones (Santerne et al., 2016). By targetting stars brighter than V~11, PLATO will detect long-period transiting exoplanets, down to Earth's size, that will be amenable to high-precision RV follow-up. The HIP41378 system is therefore a prototype of the systems that PLATO will discover, albeit it hosts relatively large planets. Characterising this system is therefore a testbed for the upcoming follow-up of PLATO's systems which will mainly rely on the ESO facilities for the southern field.

The follow-up of long-period transiting exoplanets is facing several challenges that are not seen for short-period planets. The decade-long monitoring of this system, with instruments like HARPS and ESPRESSO (Grouffal et al., 2026), allowed us to learn a few lessons that will be important in the context of the PLATO mission. We are summarising them in this paper. They are of two kinds: 1- the RV follow-up, and 2- the transit follow-up of these long-period planets.

### The HIP41378 system in a nutshell

The most up-to-date and detailed view of the HIP41378 system is available in Grouffal et al. (2026). The central star hosts five transiting planets (b, c, d, e, and f) with radius between Neptune and Saturn, and a sixth non-transiting planet (g) that has been detected by both RV (Grouffal et al., 2026) and transiting timing variations (hereafter TTVs, Leonardi et al., 2025). The system appears to be a very long chain of commensurability, with orbital periods close to 2 weeks (b), one month (c), 2 months (g), 9 months (d), one year (e), and one and half year (f). All of these planets have masses close to the one of Neptune or below, which lead to RV amplitudes between ~0.6 m/s and ~2 m/s. Grouffal et al. (2026) detected a 7th signal with an orbital period of 6 to 8 years whose planetary nature still need to be confirmed.

The three outermost transiting planets (d, e, and f) have long transit durations, from 12 to 19 hours and relatively few transits detected so far from sparse photometry. This limits the precision on their orbital periods (Becker et al., 2019 ; Berardo et al., 2019), albeit this should not be a limitation for PLATO which will continuously monitor the target stars, and hence detect consecutive transits.

The host star is a relatively quiet F-dwarf, with photometric variability at the level of 200ppm (Santerne et al., 2019) and a rotation period of 6.2 days (Grouffal et al., 2025).

## Radial-velocity follow-up of long-period planets in multi-planetary system

### Need for dense and long radial-velocity time series

Extreme-precision RV blind-search programmes are capable to detect low-mass planets at long period by targeting the most quiet stars (e.g. HD20794, Nari et al., 2025). However, photometric surveys are detecting planets transiting stars with any level of variability. Long-period planets have a low transit probability (~1% of below), and thus are extremely rare around bright and inactive solar-type stars. The RV follow-up of PLATO planets will thus have to deal with some level of stellar variability. It is relatively easy to detect low-mass exoplanets transiting stars with a moderate activity when their orbital period is shorter than the stellar rotation (e.g. Santerne et al., 2018). In the case of HIP41378, the orbital period of the outermost transiting planet (f) is about one magnitude longer than the rotation period. To overcome this limitation, the star was intensively monitored during 580 individual nights from 2016 to 2024 using mainly the HARPS, HARPS-N and ESPRESSO spectrographs (Grouffal et al., 2026) with a median timespan between two consecutive observations of about one night (see Fig 1.). This represents 111h, 51h, and 95h of HARPS, HARPS-N, and ESPRESSO time (respectively). This dense and long time series allowed Grouffal et al. (2026) to detect the RV signals of the six Neptune-mass planets with a precision from ~7% to ~50%. Without such a dense and long time series, it would not have been possible to detect and characterise this rich system.

### Commensurability between planets leads to intermittent signal

The planets in the HIP 41378 system have commensurable orbital periods. Planets d, e, and f are close to a triple-planet resonance chain (Grouffal et al, 2026). As a consequence, the RV signal of these planets appears as intermittent: some observing seasons the Keplerian orbits have oppositive variations, resulting in a nearly flat signal, while the following season, their signals add up resulting in a large variation (see Fig. 1). Moreover, although the signal of the planet candidate h is large (~1.8 m/s) and highly significant, it was only detected during the $8^{th}$ year of observation[b]. During the first seven years, its signal was absorbed by the other planetary signals, leading to their inaccurate mass measurement (see the mass difference of planet f between Santerne et al., 2019: 12±3 $M_e$ and Grouffal et al., 2026: 25±5 $M_e$). Therefore, to accurately measure the mass of a long-period planet, it first requires to detect all the other potential planets in the system whose orbital periods may be commensurable with the planet's orbital period.

These other potential planets with a commensurable period are less likely to transit if they are at long, than at short period. Indeed, if mutual inclination between planets is constant or increasing as we move away from the star, systems with short-period planets are expected to be more complete, as seen in transit, than systems with long-period ones. Moreover, if long-period planets in multi-planetary systems are commonly found in commensurability (as observed in systems with short-period planets, e.g. Lissauer et al., 2011), it means that a relatively large fraction of the transiting planets at long period that PLATO will detect will have planetary companions that are not transiting. Their RV characterisation will hence require long observing baselines to disentangle all the signals and characterise accurately the planets.

### Commensurability with the Earth's orbit

Planets with orbital periods of a few hundred days have orbital periods close to that of the Earth. It leads to strong seasonal sampling effects in ground-based observations. It means that over one observing season, it is not possible to cover their full orbit, unless the host star is close to the Earth's rotation pole. The planet HIP41378 e has an orbital period of ~390 days. Given the visibility of the star from the ground (about 6 months) the nine years of RV monitoring allowed Grouffal et al. (2026) to cover ~80% of the planet's orbit. This decrease the precision one can achieve on the mass and orbital eccentricity of the planet, especially if these planets have commensurability with others in the system. Thus, once again, the characterisation of these planets requires long observing baseline.

## Transit follow-up of long-period exoplanets

### Large transit timing variations

The amplitude of the TTVs scales with the orbital period so long-period planets might have large TTVs, at the level of hours to days (e.g. Bryant et al., 2021, Sulis et al., 2024, Leonardi et al., 2026). As a consequence, scheduling a transit follow-up of a long-period planet requires long baseline which is challenging from the ground, except for facilities located in Antarctica (e.g. Rea et al., 2025). A monitoring of long-period transit events is thus mandatory to precisely predict future transits accounting for TTVs. Without this monitoring, the difference between the true and predicted transit times could be larger than one observing night after a few orbital periods. Follow-up observations would thus be missed if TTVs are not taken into account. For instance, a transit of HIP41378 d was scheduled with the CHEOPS space mission in January 2023 only six orbits after the last transit detected by K2 (2018). The 32-h long observing run on time with linear ephemeris revealed no transit (Sulis et al., 2024). This rises doubts on the true orbital period of this planet, but these doubts are unconfirmed by recent studies (Grouffal et al., 2026 ; Leonardi et al., 2026): the planet thus exhibits TTVs of more than a day in just a few orbits.

To further explore the number of transiting planets at long period that would be missed due to large TTVs, Grouffal (2025) simulated an Earth analog (1Me, P=365d) with a non-transiting planetary companion. This analysis assumed that PLATO would have detected only two transits during its first long observing pointing in 2027 and 2028 and a follow-up observation scheduled in 2032 with e.g. ANDES / ELT. Depending on the period and mass ratios between planets, it could lead TTVs of more than 7h (hence longer than a typical observation at night) up to 87% of cases (see Fig. 2). To allow the community to follow-up long-period transiting planets detected by PLATO with future instrumentation, it is mandatory either to extend the duration of the long-duration observation phase beyond two years, to detect more than two transits, or closely monitor each future transit (after the PLATO observation ends) and model TTVs precisely. However, for planets smaller than Neptune (depth < 1mmag), this follow-up might be impossible from the ground and require space-based facilities.

## Rare opportunities to observe the long-period transits

Long-period planets offer rare opportunities to observe their transit. From the ground, each transit might not be visible, because the host star has a low visibility, or the transit occurs during daytime at a given observatory. This leads that transits might be observed only once every few years (see Fig. 3). Due to their commensurability with the Earth's orbit, the long-period planets HIP41378 d, and f could be observed from the ground with a high visibility only once every ~3 years (see Fig. 3). Unlike HIP41378 (DEC=+10º), the PLATO long-duration observation phase South #2 (LOPS2 ; Nascimbeni et al., 2025) is high in the ecliptic plane, so the visibility of the latter targets from the ground should be much better than for the former one. Nevertheless, given the opportunities to observe a transit of a long-period planet are rare, it is important to have redundancies in the facilities. As an exemple, the 2019's transit of HIP41378 f was scheduled to be observed by a dozen ground-based facilities, including four spectrographs (SOPHIE/OHP, ESPRESSO/ESO, EXPRES/DCT, PARAS/MountAbu) and about ten photometric telescopes. Given the bad weather, the technical issues, and the human mistakes, only a few observatories secured data: observations from La Palma (Spain) reported no transit detection due to unexpected large TTVs and only the NGTS observatory detected the transit (Bryant et al., 2021). This first ground-based transit allowed us to predict the following ones which were all observed since then and further characterise the planet (e.g. Alam et al., 2022 ; Grouffal et al., 2025). Transit follow-up of long-period planets should thus be observed by several facilities to secure its detection.

## The challenges of observing long transit duration from the ground

Long-period planets have long-duration transits, some of them last longer than one night which challenge ground-based observations. For instance, the 2022's observing campaign on HIP41378 f, as reported in Grouffal et al. (2025), used nine high-precision spectrographs spread across the world to cover ~70% of its 19-h long transit in a single event. This campaign was a success as HIP41378 is located near the ecliptic, hence it can be observed from most observatories in the world, while PLATO targets will be much further in the South. As a consequence, they will have a low visibility from Northern facilities. Ground-based facilities located in the Southern hemisphere are mainly located in Chile with relatively few observatories in South Africa and Australia (especially RV spectrographs). Therefore, to cover a long-duration transit like HIP41378 f (as in Grouffal et al., 2025) only from Chile, it would require to observe several events, hence spread over many years[c].

This situation is even worse in photometry as the signal varies mainly during the relatively sharp ingress or egress. For a planet like HIP41378 f, a given observatory might see an ingress or egress only once every a few transits (García-Mejía, et al., 2024 ; Leonardi et al., 2026), hence once in many years. As a consequence, monitoring TTVs of a PLATO long-period planet from the ground will require facilities spread in longitude and not only in Chile. An alternative of that would be to use the spectroscopic Rossiter-McLaughlin (RM) effect as mean to constrain transit times. Following Gaudi & Winn (2007), the RM effect might have a relatively large RV amplitude, even for small planets. Therefore, from the ground, a transiting planet might be easier to detect in spectroscopy than in photometry. Moreover, the RM effect is a continuous variation along the transit and not only during ingress and egress. In the case of HIP41378 d, which transit depth is only of 650ppm for a 12h-long event, Grouffal et al. (2022) derived a timing with a 5-min and 30-min precision for the 2019 and 2022's events (respectively). This precision might be enough for these long-period planets which exhibit TTVs at the level of hours to days (e.g. Sulis et al., 2024).

## Summary and conclusions

PLATO will be the first space mission capable of detecting planets at long orbital periods (P > 100 days) transiting bright stars that are amenable to high-precision RV follow-up observations. The first stellar field will be in the Southern hemisphere, hence the ESO instruments will play a key role in the scientific exploitation of the mission. In this paper, we listed the lessons learned from a decade of studying the HIP41378 multi-planetary system at long period which is a testbed for the forthcoming follow-up of PLATO planets. The results recently published in Grouffal et al. (2026) shows that it is possible to characterise low-mass exoplanets at long period, but this required a relatively long and dense RV time series. This time series, which cost a total of about 250h of HARPS, HARPS-N, and ESPRESSO time, was needed in order to smooth out stellar variability, disentangle the signals from the various planets in commensurability, and cover the phase of planets which orbital period is close to 1 year. The follow-up of PLATO targets will face similar challenges which are not affecting systems with short-period planets (P < 100 days). This is why a strong synergetic partnership between ESO and the PLATO consortium is foreseen in order to insure an optimum outcome of the mission.

Beyond the mass characterisation, the follow-up of long-period and long-duration transits from the ground is also a challenge. These planets might have TTVs at the level of hours to days which can strongly limit the scheduling of follow-up observations. This limitation can be fixed by extending the duration of the PLATO observation as long as possible or follow-up each transit in order to model TTVs and precisely predict future events. Long-period transiting planets have rare transit opportunities, especially from a given ground-based observatory. Each event should thus be considered as highly time critical. Their transit lasts relatively long, and could be longer than one night. Hence, covering most of the transit requires either a worldwide collaboration (as in Grouffal et al., 2025), or to observe several events with the same instrument, which would take many years to complete.

Long-period transiting planets are as much interesting as they are rare (Ulmer-Moll et al., 2026) and PLATO will probe them around bright star as never before. However, their characterisation through RV or transit follow-up is much more challenging than for short-period planets. By studying the HIP41378 system over the past decade, we are now prepared for the follow-up of multi-planetary systems at long period that will be detected by PLATO.

## Acknowledgements

This work was supported by the "Programme National de Planétologie" (PNP) of CNRS/INSU co-funded by CNES. This work was supported by the "Action Thématique Physique Stellaire" of CNRS/INSU PN ASTRO co-funded by CEA and CNES.

## Notes

a A few transiting giant planets with orbital periods beyond 100 days were also detected by the TESS satellite (e.g. Mireles et al., 2023)

b The signal from the planet candidate HIP41378 h could have been detected during the 5th observation year, but no data were collected that year due to the COVID pandemic.

c The PLATO targets have a much better visibility along the year from the ESO observatories than HIP41378 so a given event can be much better covered than in the campaign reported in Grouffal et al. (2025). Still, a 19h-long transit like HIP41378 f would require several events to cover from Chile, which would take up to one decade depending on the orbital period, transit duration and visibility of the target.

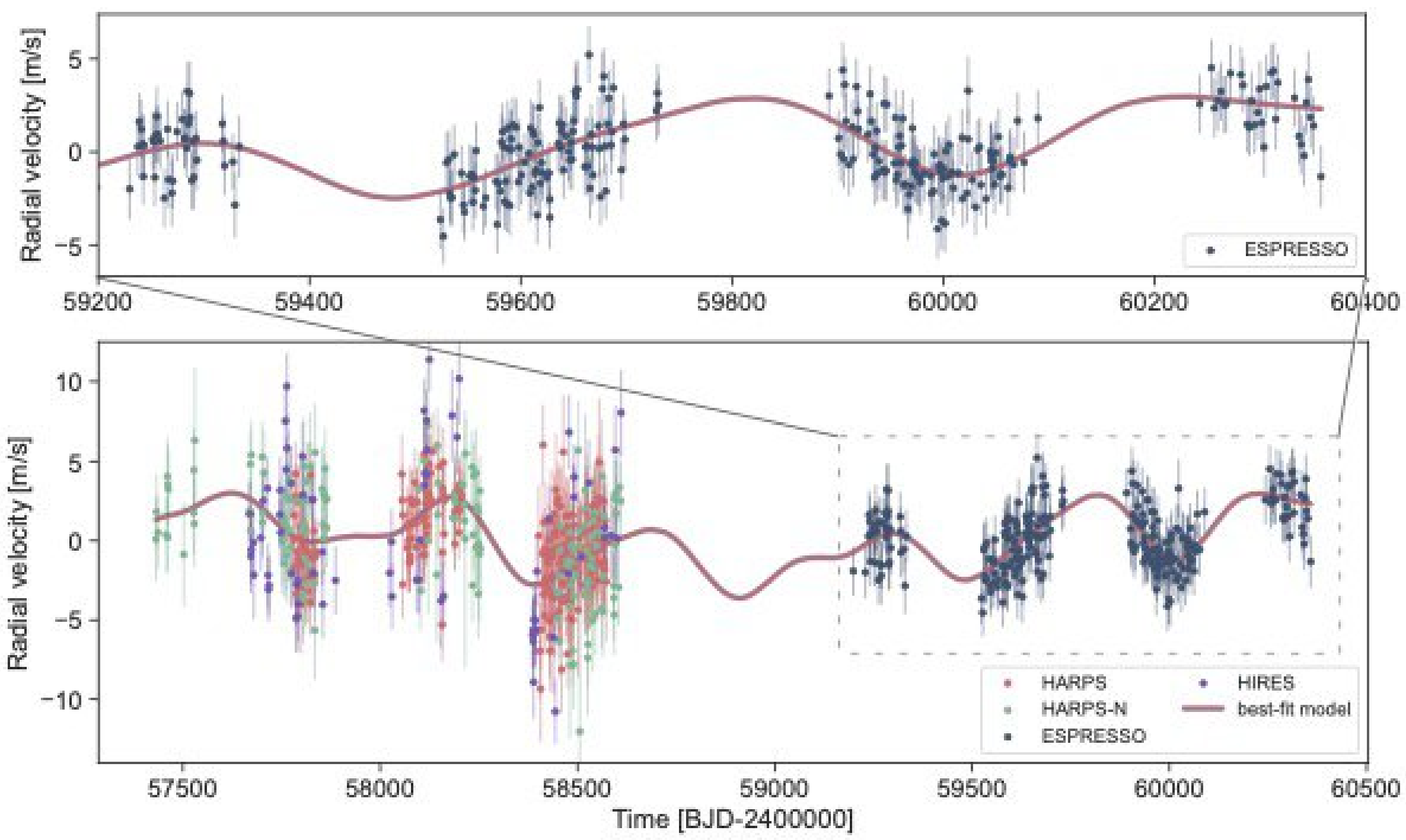

**Figure 1.**
The nine year RV monitoring of HIP41378 with HARPS, HARPS-N, HIRES, and ESPRESSO spectrographs. The star was nightly observed whenever possible, except in 2020 during the COVID pandemic. The high-frequency signals (planets b, c, g, and the stellar rotation) have been removed from the data and model to highlight the long-period signals. Adopted from Grouffal et al. (2026).

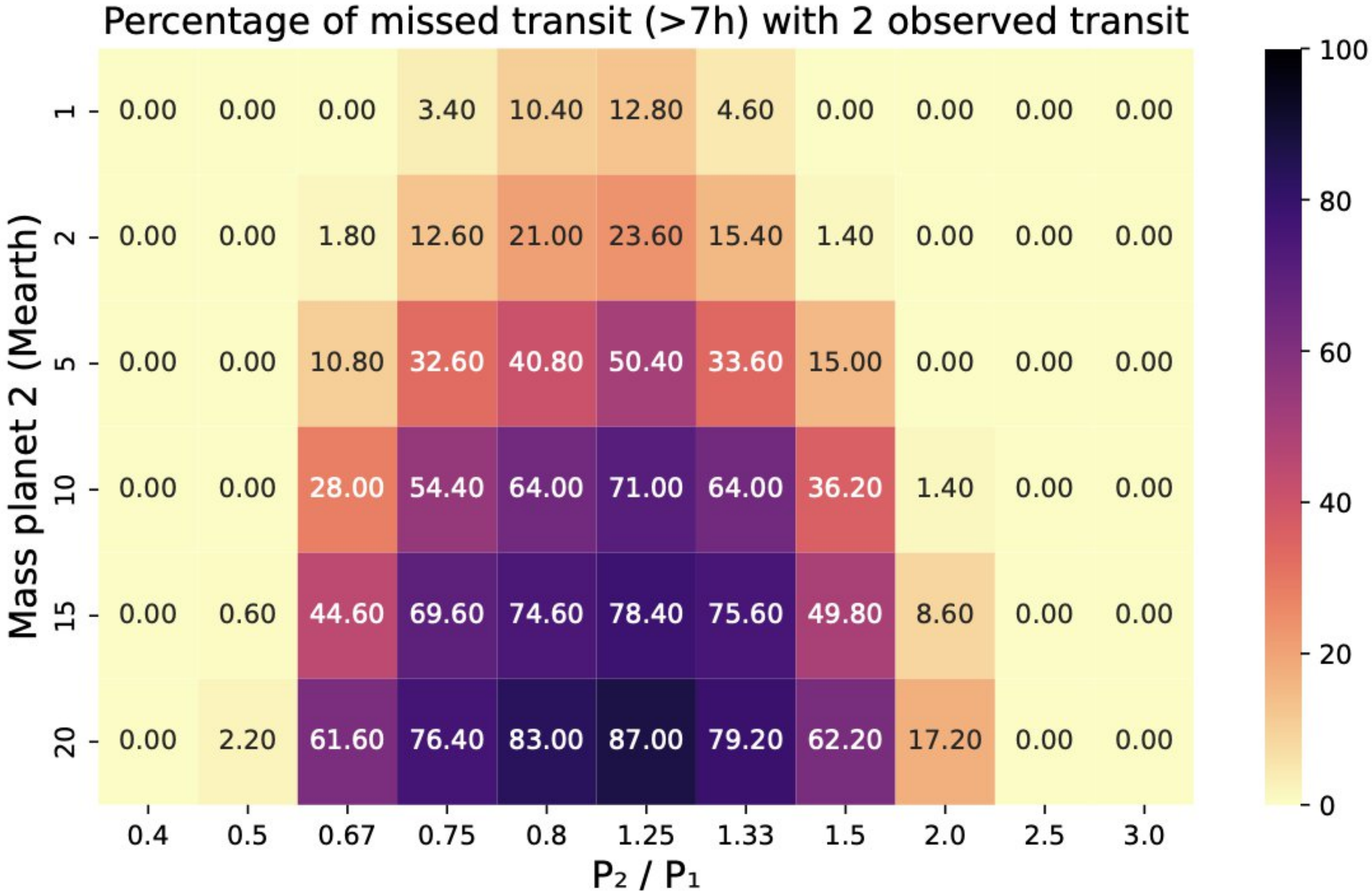

**Figure 2.**

Assuming an Earth analog (1Me, P=365d) seen twice in transit by PLATO in 2027 and 2028, this is the ratio of transits that would be missed, i.e. their TTV is longer than one night of 7h, when scheduling a follow-up observation with, e.g. ANDES/ELT, in 2032. This ratio depends on the mass and period of the (unseen in transit) perturber. In the most extreme scenarios, more than 80% of the transiting planets observed only twice by PLATO could not be re-observed a few years later. Figure from Grouffal (2025).

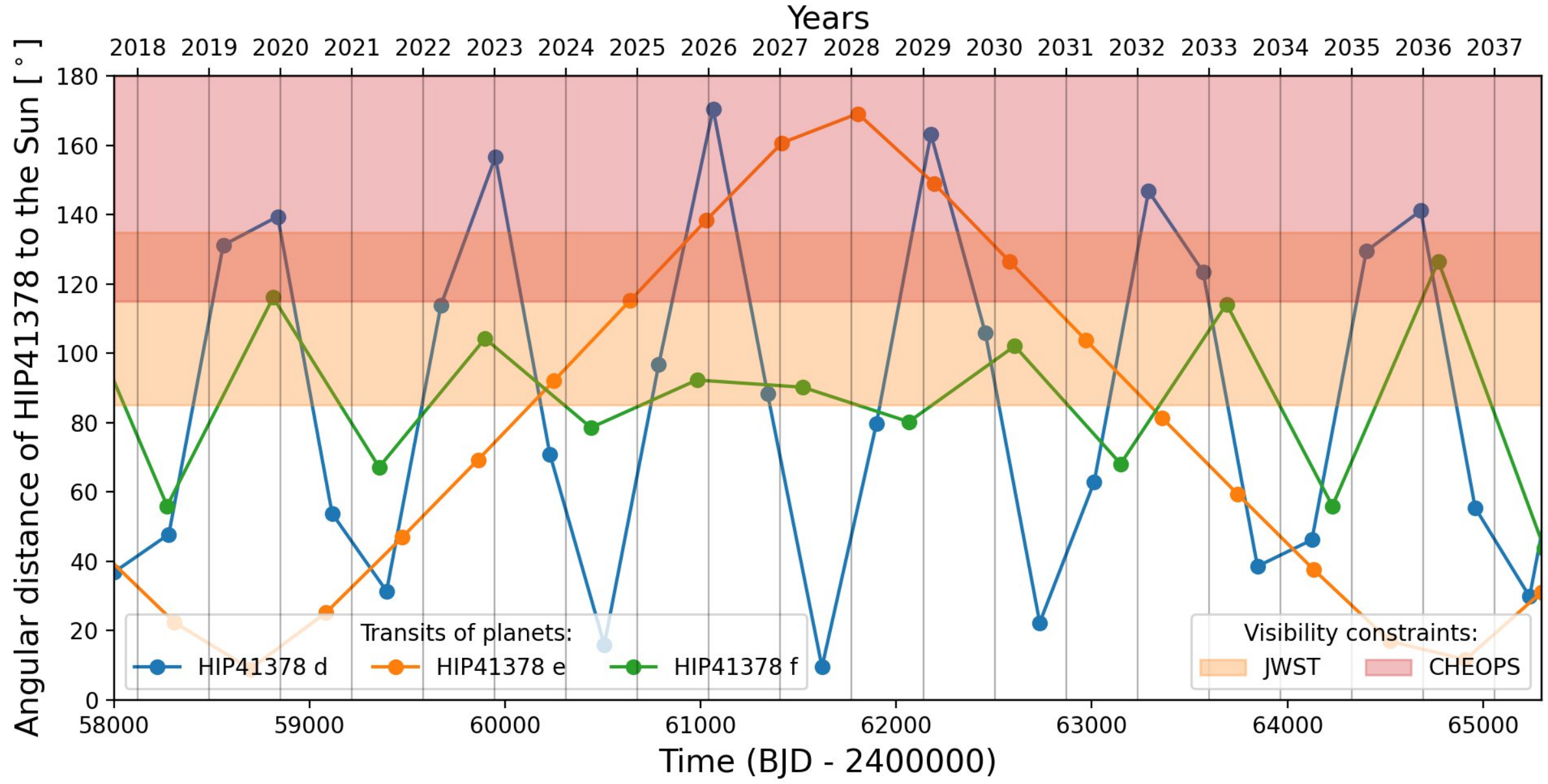

**Figure 3.**
Angular distance of the star HIP41378 to the Sun, as seen from the Earth, at the expected times of transit (dots) of the planets d (blue), e (orange) and f (green). The visibility constraints from JWST and CHEOPS are displayed in yellow and red (respectively). The period covers 20 years of time, from the K2's campaign C18, to the first light of ELT/ANDES (expected after 2032). Since the planets have orbital periods in commensurability with the Earth, there are only a few transit events with high visibility per decade, and even less for HIP41378 e. Some of these high-visibility events are transiting during the day or when the star is down from a given observatory.